\documentstyle[preprint,eqsecnum,aps,epsf]{revtex}

\newcommand{\bi}{\bibitem}
\newcommand{\be}{\begin{eqnarray}}
\newcommand{\ee}{\end{eqnarray}}
\newcommand{\nn}{\nonumber}
\catcode`\@=11
\def\lsim{\mathrel{\mathpalette\@versim<}}
\def\gsim{\mathrel{\mathpalette\@versim>}}
\def\@versim#1#2{\vcenter{\offinterlineskip
\ialign{$\m@th#1\hfil##\hfil$\crcr#2\crcr\sim\crcr } }}
\catcode`\@=12

\begin{document}
    
\tightenlines

\draft
\title{Unrenormalizable Theories Can Be Predictive.}
\author{Jisuke Kubo$\ ^{(a,b)}$ and Masanori Nunami$\ ^{(b)}$}
\address{
$\ ^{(a)}$ 
Max-Planck-Institut f\"ur Physik,
 Werner-Heisenberg-Institut \\
D-80805 Munich, Germany\\
$\ ^{(b)}$ Institute for Theoretical Physics, 
Kanazawa  University, 
Kanazawa 920-1192, Japan
}
\maketitle
\begin{abstract}
Unrenormalizable theories contain
infinitely many free parameters.
Considering these theories
in terms of the Wilsonian renormalization group (RG),
we suggest a method for
removing  this large ambiguity.
Our basic assumption is the existence
of the maximal ultraviolet  cutoff
in a cutoff theory,  and 
we require that 
the theory be so fine-tuned as to
reach the maximal cutoff.
The theory so obtained behaves as a local
continuum theory to the  shortest distance.
In concrete examples of the scalar theory we find that 
at least in a certain approximation to
the Wilsonian RG, this requirement enable us
to make unique
predictions in the infrared regime
in terms of a finite number of independent parameters.
Therefore,  the method
might provide a way for calculating quantum corrections
in a low-energy effective theory of quantum gravity.

\end{abstract}
\vspace{1cm}


\narrowtext

\section{Introduction}
Quantum field theories are classified according to their renormalizability.
Needless to say that  not only renormalizable, but also unrenormalizable
 theories have
played important r\^ ole in particle physics \cite{weinberg1}.
It is however widely accepted that  an unrenormalizable
 theory is only a  low-energy
effective theory of a more fundamental theory.
The low-energy effective theory should contain the correct
low-energy degrees of freedom 
of the fundamental theory, and 
it should be
possible, within the framework of
the effective theory,  to 
compute  approximate low-energy {\em quantum} processes
of the fundamental theory.

If the effective theory is perturbatively
unrenormalizable, 
we face a serious problem.
How many independent parameters 
should have the perturbatively unrenormalizable theory?
The answer in perturbation theory is: Infinitely many, by definition. 
Of course, this does not prevent from
applying perturbation theory to unrenormalizable theories
to make predictions, as in  chiral
perturbation theory which has a certain success 
\cite{chiralpt,gomis,ecker}.
Nevertheless we may ask why chiral
perturbation theory has so many independent parameters 
at the quantum level,  although
it is the effective theory of QCD. 
 This large ambiguity  cannot be controlled 
 by a symmetry \cite{chiralpt,gomis},
and we are concerned with this problem
which always exists in unrenormalizable effective theories.
In this paper, we propose to remove this large
ambiguity within the framework of
the effective theory.

The idea is based on a simple  intuitive picture.
Suppose we formulate both a  fundamental theory and
its low-energy effective theory 
 in terms of the Wilsonian renormalization group (RG) \cite{wilson1}. 
We assume that the effective theory is 
a cutoff theory.
Since we assume that the fundamental theory is free from the
ultraviolet cutoff, $\Lambda_0$,
we can let go $\Lambda_0$ to infinity.
That is, starting at
some point in the infrared regime,
the RG flow  in the fundamental theory evolves along 
a renormalized trajectory, and
approaches an ultraviolet fixed point in the ultraviolet limit.
The flow has to evolve  for ``infinite time'' to
arrive at the fixed point \cite{wilson1}.
The RG flow in the effective theory
that  evolves for the `` maximal time''
may be  the best approximation to the renormalized trajectory
of the fundamental theory,
and the effective theory along that trajectory
behaves as a local continuum theory down to
 the shortest distance.
 The large ambiguity of the cutoff theory could be removed in 
 this way.  In Sec. II we will formulate our idea.

We will consider
the scalar theory in four, five and six dimensions in Sec. III, and 
apply our idea to these theories.
It will be shown that 
in lower orders in our approximation
to the Wilsonian RG, the  ambiguities inherent in
these  theories can be removed.
We will  also argue that the ambiguity
in four dimensions, which we will fix,
is  the renormalon ambiguity \cite{renormalon,luscher1}.
Einstein's theory of gravity and also Yang-Mills theories in more than 
four dimensions 
are perturbatively unrenormalizable, and
the conventional method of perturbation theory
loses its power in these theories \footnote{See Ref.~\cite{carlip}
for recent progress in quantum gravity.}.
The application to
quantum  gravity and also 
higher-dimensional Yang-Mills theories would go beyond the scope
of this paper, and we leave it to feature work.
However, as far as renormalization is concerned,
the scalar theory in 
more than four dimensions may be seen as an oversimplified model of 
a low-energy effective theory of quantum 
gravity \cite{weinberg3} \footnote{
Our approach is related to  Ref.~\cite{souma}.}.

We conclude in Sec. IV, and the explicit expressions for
the $\beta$ functions which we use 
in an approximation scheme that is specified 
in Sec. III are given in Appendix.

\section{Making unrenormalizable theories predictive.}

Our interest is directed to field theories \cite{wilson1} which become weakly
coupling in the infrared regime.
These theories can be 
perturbatively renormalizable or  unrenormalizable. 
Suppose we consider such a theory 
in terms of the  Wilsonian RG \cite{wilson1}. 
We  then gradually increase  the ultraviolet cutoff $\Lambda_0$
while keeping fixed the values of  the coupling constants of the theory
at some point $\Lambda$ in the infrared regime, and
consider the RG flow
as a function of $\Lambda_0$.
In the case of  a nontrivial theory, 
the ultraviolet cutoff $\Lambda_0$ can become infinite,
and the RG flow converges to a fixed point,
if a certain set of the coupling constants is exactly fine tuned
at $\Lambda$, that is, if they lie exactly on
a renormalized trajectory \cite{wilson1}.
If the theory is trivial, $\Lambda_0$ cannot become infinite,
or the RG flow does not converge to a point.
Suppose there
exists the maximal value $\Lambda_{\rm max}$ in the theory.
To reach $\Lambda_{\rm max}$, we have to fine tune
the values of the coupling constants at $\Lambda$
as in the case of a nontrivial theory.

With this observation, we now come to formulate 
our method under the basic assumption that there
exists the maximal value 
of the ultraviolet cutoff $\Lambda_{\rm max}$ in a theory.
If the theory is
 perturbatively renormalizable, there should be
a set of dimensionless coupling constants.
In this case we regard all the coupling constants with a canonical 
dimension $< 0$ as dependent coupling constants.
If the theory is
perturbatively unrenormalizable, there should exist a set of coupling
constants with a negative  canonical dimension $\geq d_{\rm max}$,
which should be regarded as independent.
In this case we regard all the coupling constants with a canonical 
dimension $ < d_{\rm max} $ as dependent coupling constants.
With this classification of
the independent and dependent coupling constants,
we then require from
the dependent coupling constants that 
for  given values of the independent coupling constants in the 
infrared regime, 
the dependent ones are so fine tuned that one arrives at
the maximal value of the ultraviolet cutoff
\footnote{This idea is in fact similar to the principle of minimal 
sensitivity \cite{stevenson}
which tries to resolve
the renormalization scheme ambiguity in perturbation theory.}.

Polchinski \cite{polchinski}
investigated the Wilsonian RG flow 
of the scalar theory in four dimensions  to show
its perturbative renormalizability 
\footnote{
For extension to Yang-Mills theories, 
see \cite{becchi}, for instance.}.
His observation, assuming that the mass parameter
can be neglected, is that there exists a  trajectory
in the space of coupling constants, which is
attractive in the infrared limit. 
That is, whatever the initial values of the coupling constants at
$\Lambda_0$ are, they converge to the trajectory,
if the $\Lambda_0$ is large enough.
This is then interpreted as perturbative renormalizability.
Therefore, in the class of theories, in which
Polchinski's criterion on perturbative renormalizability
can be applied, the ultraviolet cutoff
$\Lambda_0$ should be sufficiently large
so that physics in the infrared regime
has less dependence of 
the initial values of the coupling constants at $\Lambda_0$.
It should be therefore assumed  for  perturbative renormalization
to work  in those trivial,
but perturbatively renormalizable theories
that {\em the intrinsic cutoff of the theory is so large
that the nonperturbative effects due to
the finite intrinsic cutoff
are very small in the infrared regime} \footnote{See for instance
Ref. \cite{luscher1}.}. 
If the intrinsic cutoff is low, the initial value
ambiguities are
not  sufficiently suppressed in the 
infrared regime, and consequently the perturbative calculations
may not  be reliable.
Therefore, the intrinsic cutoff can be much larger than
the actual physical cutoff at which a new physics enters.
As for these perturbative renormalizable theories,
{\em our requirement  is a slight extension of the 
requirement in perturbative renormalization;
the maximal, instead of a large, cutoff is assumed.}

According to Polchinski \cite{polchinski},
we interpret  a theory as perturbatively unrenormalizable,
if there is no infrared attractive trajectory
(or no infrared attractive subspace of a finite dimension).
Physics in the infrared regime in this case depends
on the initial values of the infinitely many coupling
constants at $\Lambda_0$.
This is exactly the situation  which we are more interested in.
Although in the case of
perturbative renormalizable theories,
our requirement  of the maximal
cutoff is a slight extension,
this requirement,  applied to a  perturbatively unrenormalizable
theory, could   select a trajectory out the
infinitely many trajectories.
The theory along this trajectory will behave as a local continuum theory
to the shortest distance.
In the next section we will consider
concrete unrenormalizable theories, and  see that our idea works 
at least in a certain approximation.
If our idea could be founded in a more rigorous manner,
it could be promoted to a principle, which we would like to call
principle of maximal locality.
It is, however, beyond the scope of the present paper
to do this task.

\section{Application  to the scalar theory\\
    in diverse dimensions}

\subsection{Continuous Wilsonian renormalization group }
As we have explained in detail in
Sec. II, our interest is directed to trivial theories.
To define such theories in a nonperturbative fashion,
we have to introduce a cutoff.  
A natural framework to study 
cutoff theories is provided by the continuous Wilsonian RG 
\cite{wilson1}\cite{wegner}--\cite{legendre}.
Let us briefly  illustrate the basic idea of the Wilsonian RG approach
in the case of the $N$ component scalar 
theory in euclidean $d$ dimensions \footnote{See 
for instance Ref.~\cite{aoki2}
for review.}.
One  divides the field  $\phi(p)$ in the momentum space into
low and high energy modes according to
\be
\phi^k (p)=\theta(|p|-\Lambda)~\phi_{>}^k(p)+
\theta (\Lambda-|p|)~\phi_{<}^k (p)~, ~k=1,\dots,N~.
\label{modes}
\ee
The Wilsonian effective action is then defined by 
integrating out only the high energy modes in the path integral:
\be
S_{\rm eff} [~\phi_{<},\Lambda~]
=-\ln \left\{
\int {\cal D}\phi_{>} ~e^{-S[\phi_{>},\phi_{<}]}
\right\}.
\label{wilson}
\ee
It was shown in Refs.~\cite{wilson1,wegner} that the path integral
corresponding to the difference
\be
\delta S_{\rm eff} =
S_{\rm eff} [~\phi_{<},\Lambda+\delta\Lambda~] - S_{\rm eff} 
[~\phi_{<},\Lambda~]
\ee
for an infinitesimal $\delta\Lambda$ can be exactly carried out,
yielding  the RG evolution equation of the effective action
\be
\frac{\partial S_{\rm eff}}{\partial t} &=&
-\Lambda \frac{\partial S_{\rm eff}}{\partial \Lambda}
= {\cal O}(S_{\rm eff})~,
\label{nrg1}
\ee
where ${\cal O}$ is a non-linear operator acting on the functional 
$S_{\rm eff}$. 
There exist various (equivalent) formulations 
\cite{wegner}--\cite{legendre}, but in this paper we
consider only the Wegner-Houghton  (WH) equation
\cite{wegner}.
Since $S_{\rm eff}$ is a functional of fields, one can 
think of the  WH equation as coupled differential
equations for  infinitely many couplings in the effective action.
The crucial point is that ${\cal O}$ can be exactly derived for a
given theory, in contrast to the perturbative RG approach where the 
RG equations are known only up to a certain order in perturbation 
theory. This provides us with possibilities to use approximation
methods that go beyond the conventional perturbation theory.

In the derivative expansion 
approximation 
\cite{nicol,hasenfratz}-\cite{wetterich2}-\cite{aoki1},
one assumes that the effective action $S_{\rm eff}[\phi,\Lambda ]$ 
can be written as a space-time integral of a (quasi) local function
of $\phi$, i.e.,
\be
S_{\rm eff}[\phi,\Lambda ] &=&
\int d^d x\,(\,\frac{1}{2}\, \sum_{k,l=1}^N
\partial_{\mu}\phi^k\partial_{\mu}\phi^l\,Z^{kl}(\phi,\Lambda)
+V(\phi,\Lambda)+\dots~)~,
\label{ansatz1}
\ee
where $\dots$ stands for terms with higher order derivatives
with respect to the space-time coordinates.
In the lowest order of the derivative expansion (the local potential
approximation \cite{nicol,hasenfratz}), 
there is no wave function renormalization 
($Z^{kl}(\phi,\Lambda)=\delta^{kl}$),
and the RG equation for the effective potential
$V$ can be obtained. 
Since it is more convenient to work with the RG equation for
dimensionless quantities, which makes the  scaling properties
more transparent, 
we rescale the quantities according to
\be
p &\to \Lambda p~,~\phi^k \to \Lambda^{d/2-1} \phi^k~,~
V \to \Lambda^{d} V~.
\label{scaling1}
\ee
Then the RG equation $V(\phi,\Lambda)$  is given by \cite{hasenfratz}
\be
\frac{\partial V}{\partial t} & =& 
-\Lambda\frac{\partial V}{\partial \Lambda}= a~
\ln (1+V'+2 \rho V^{\prime\prime})
+a (N-1)\ln (1+V')+d V+(2-d) \rho V' ~,
\label{wh}
\ee
where the prime on $V$ stands for the derivative with respect to 
$\rho$, and
\be
\rho &=&  \frac{1}{2}\sum_{k=1}^N \phi^k \phi^k ~,~
a= \frac{1}{2^d\pi^{d/2} \Gamma(d/2)} ~.
\label{ad}
\ee
Eq.  (\ref{wh}) ( or the one which is derived from (\ref{wh}) for
$F=V'=\partial V/\partial \rho$)
is the central equation we will analyze 
in the following  subsections.
Therefore,  all the results we will obtain are valid only within the
local potential approximation.
There are however no fundamental problems 
in going beyond this approximation.

\subsection{Toy model in $d=3$}
The scalar theory in $d=3$ dimensions have a nontrivial fixed point,
Wilson-Fisher fixed point \cite{wilson1}, 
and is moreover asymptotically free.
So this theory is nontrivial as it is well known, and therefore lies
outside of our interest. However, we would like to consider
this theory with infrared and ultraviolet interchanged in order to
illustrate what we mean by `` being
closest to a nontrivial theory''.
So, the results obtained here will be compared with those
of trivial theories which we will consider later.

The interchange of infrared and ultraviolet just means 
the replacement
$t \to -t$.
We then consider $F=V'=\partial V/\partial \rho$ for $N=1$, 
and derive the
evolution equation  from (\ref{wh}) for $F$
\be
\frac{\partial F}{\partial t} &=&
-2 F+ \rho F'
-a ~\frac{3 F'+2 \rho F^{\prime\prime}}{1+ F+2 \rho F'}~,~
a=\frac{1}{4\pi^2}~.
\label{rged3}
\ee
The power series ansatz
\be
F(\rho,t)&=&4a ~f_0(t) +\sum_{n=1}^{\infty} f_n(t) ~[\frac{\rho}{4a}]^{n}
\ee
defines the coupling constants $f_n$, and Eq. (\ref{rged3}) gives a set of 
$\beta$ functions which in lower orders in the expansion are given by
\be
\beta_0 &=&\frac{d f_0}{d t}=-2 f_0-\frac{3}{4}
\frac{f_1}{(1+f_0)}~,\\
\beta_1 &=& \frac{d f_1}{d t}=-f_1+\frac{9}{4}
\frac{f_1^2}{(1+f_0)^2}~,
\ee
Where we have made a truncation at $n=1$ above. 
There exist two fixed points
\be
(f_0= 0, f_1=0)~, ~(f_0= -\frac{1}{7}, f_1=\frac{16}{49} )~.
\label{fpd3}
\ee
All the directions from the 
Gaussian fixed point (at this level of truncation) are infrared stable;
the eigenvalues are
$(-1,-2)$ with the corresponding eigenvectors 
$(-3/4,1)$ and $(1,0)$, respectively.
The solution near the origin can easily be found:
\be
f_0 &=& -\frac{3}{4}f_1 +[~C
-\frac{27}{16}\log(f_1)~] f_1^2+O(f_1^3)~,
\label{sold3}
\ee
where $C$ is an integration constant.
The solution is obtained from $
d f_0/d f_1=\beta_0/\beta_1$.
At this stage, $f_0$ and $f_1$ behave independently.
Note that  there is a certain infrared attractiveness,
because $f_0$ approaches to a definite function,
$  -(3/4)f_1 $, in the infrared limit $f_1 \to 0$.

The integration constant $C$ can be determined  by requiring
that the RG flow approaches to the ultraviolet fixed point
$(f_0= -1/7, f_1=16/49)$.
That is, the coupling constant $f_0$ has to be
exactly fine tuned for the ultraviolet cutoff $\Lambda_0$
to become infinite.
If $f_0$ is not exactly fine tuned, the RG flow runs
into infinity at some finite $\Lambda_0$.
The fine-tuning procedure is shown in Fig.~1 for
$f_1=0.01$ at $t=0$,
where the vertical axis is
$T_0=\log\Lambda_0/\Lambda$ 
and the horizontal one is $f_0(0)$. 
From Fig.~1 we find that in this case
$f_0$ at $t=0$ should be fine tuned  at $-0.00718\ldots$.

\subsection{$d=4$: Perturbatively renormalizable case}
We now come to discuss a  trivial, but
perturbatively renormalizable case
\footnote{We assume that the scalar theory in four dimensions is
trivial \cite{trivial1,luscher1,luscher2}.}.
In this case, our
interest focuses on the nonperturbative ambiguity 
that results
from the fact that perturbation series diverge.
There are rigorous results that 
this divergence is dominated by the renormalon singularity
in the Borel plane, and 
the form of the ambiguity is known \cite{renormalon}.
Since the Wilsonian  RG is nonperturbative, it is natural 
to assume that it also contains this information. 
Of course, it is not clear that
one can see it  within the framework of our approximation.
We will argue, based on our 
numerical analysis,
 that this is the case.

As in the case for $d=3$, we consider the derivative
$F=\partial V/\partial \rho=V'$, where the potential $V$
is assumed to be expandable 
 as
\be
V(\rho,t) &=& v_0(t)+\sum_{n=1}^{\infty}
\frac{1}{n+1}\frac{f_n(t)}{(4 a)^n} ~
[~\rho-4af_0(t)~]^{n+1}~
\label{ansatz}\\
& =& v_0(t)+\frac{1}{2}\frac{f_1(t)}{4 a}
[~\frac{1}{2}{\bf \phi}\cdot {\bf \phi}-4 a f_0(t)~]^2
+\frac{1}{3}\frac{f_2(t)}{(4 a)^2}
[~\frac{1}{2}{\bf \phi}\cdot {\bf \phi}-
4 a f_0(t)~]^3+\cdots~.\nn
\ee
The constant $a$ is given in (\ref{wh}) ($a(d=4)=1/16\pi^2$), and
$\rho={\bf \phi}\cdot {\bf \phi}/2=
\sum_{i=1}^N \phi_i \phi_i/2$.
Here we have shifted $\rho$ by $4a f_0(t)$ from the reason we
will give soon\footnote{Shifting
$\rho$  improves
the truncation dependence in computing critical exponents in lower dimensions
\cite{morris,aoki1}.}.
The squared mass is $2 f_0 f_1$ so that if 
$\lim_{t \to \infty } 2 f_0(t) f_1(t) ~> (<) ~0$, we are in the
(un)broken phase, and the critical surface is defined by 
the RG flows that satisfy $\lim_{t \to \infty } 2 f_0(t) f_1(t) =0$.
Inserting the power series ansatz 
(\ref{ansatz}) into the evolution equation for $F$
\be
\frac{\partial F}{\partial t} &=&
2 F+ (2-d) \rho F'
+a \left[~\frac{3 F'+2 \rho F^{\prime\prime}}{1+ F+2 \rho F'}
+(N-1) \frac{F'}{1+F}~\right]~,
\label{rged4}
\ee
we can obtain the $\beta$ functions,
$\beta_n = d f_n/d t$, at any finite order of truncation.
(The explicit expressions in lower orders are given
in Appendix A.)
One can convince oneself that the $n$th order $\beta$
function has the form
\be
\beta_n& =& (2+2n-n d) f_n+
\sum_{i_1+2 i_2+\cdots+n i_n=n+1}~\eta_{ i_1 \cdots i_n}
~f_1^{i_1}\cdots f_{n}^{i_{n}}\nn\\
& &
+\sum_{l=1}^{n+1} \sum_{m=0}^{n+1}\left[
\sum_{-i_{l0} +\cdots  +m i_{lm}=n+1}~\chi^{(l)}_{
i_{l0} \cdots i_{lm}}
~f_0^{i_{l0}} f_1^{i_{l1}}\cdots f_{m}^{i_{lm}}~\Delta^{l} \right]~,
\label{bn}
\ee
where
\be
\Delta=(1+2 f_0 f_1)^{-1}~.
\label{delta}
\ee

Given the $\beta$ functions, we investigate 
the infrared and ultraviolet behavior of the theory, and
we first investigate
 the infrared behavior for
 \footnote{We have chosen the case for $N=4$, because
 we would like to apply the results below
 to the standard model elsewhere.}
 \be
 d&=&4~~\mbox{and} ~~N=4~.
 \ee
Obviously,
($f_0=3/4~,~f_n=0 ~(n \geq 1)$)
is a fixed point (Gaussian fixed point).
The stability of the RG  flows near the Gaussian fixed point 
cannot be simply 
discussed in the present case, because the $\beta$ functions are singular at this
point ($1/f_1$ which is present in $\beta_0$ appears in 
other $\beta$ functions).  So
we construct explicitly the solution near the Gaussian fixed point
to investigate its stability. To this end, we eliminate
$t$ in favor of $f_1$ using the set of equations
\be
\beta_1 \frac{d f_n}{d f_1} & =& \beta_n~,~(n\neq 1)~,
\label{reduction}
\ee
which is called the reduction equation 
in Ref.~\cite{reduction}.
We find that the power series solution 
\be
f_0 &=&   \frac{3}{4}+  \sum_{l=0}^{\infty}C^{(l)}_{0} f_1^{l+1}~,
\label{power0}\\
f_n &=& f_1^{n+1}\sum_{l=0}^{\infty} C^{(l)}_{n} f_1^l~
\label{powern}
\ee
exists. That is, the expansion coefficients can be uniquely
computed for a given truncation of the series (\ref{ansatz}).
Using this power series solution, we obtain the general solutions
that surround this special solution by solving
the linearized equations
\be
-3 f_1^2 \frac{d \delta f_0}{d f_1} &\simeq &
2\delta f_0-\frac{3}{4 f_1}\delta f_2~,\nn\\
-3 f_1^2 \frac{d \delta f_n}{d f_1} &\simeq &
[~(2-2 n) -3 p_n f_1~]\delta f_n+\sum_{i\neq n}q_(f_1) \delta f_i~.
\label{linear0}
\ee
Therefore, the integration constant $K_{n} $
of $\delta f_n$ appears in the exponential form
\be
K_{n} \exp (\frac{2-2n}{3f_1})~f_1^{p_n}~,
\label{deltafn}
\ee
where 
$p_n$ are fractional numbers.
We find that for $n \leq 2$
\be
f_0 &=&
\frac{3}{4 }- \frac{9}{16}f_1^2 + \frac{225}{64} f_1^2 - 
\frac{7857}{256}f_1^3 + 
\frac{269001}{1024 }f_1^4- \frac{12806991}{4096}f_1^5 + 
 \frac{650870883}{16384}f_1^6+O(f_1^7)\nn\\
 & &
+ K_2\exp (-\frac{2}{3f_1}- \frac{57}{4}f_1)~
f_1^{7/2}~[\frac{3}{16} - \frac{9}{8 } f_1 
+ \frac{2799}{256}f_1^2+O(f_1^3)~]~,
\label{general0}\\
f_2 &=& 
\frac{15}{4} f_1^3 - \frac{189}{8 }f_1^4+ 
\frac{7479}{64 }f_1^5- \frac{12879}{32 }f_1^6+O(f_1^7)
+ K_2\exp (-\frac{2}{3f_1}- \frac{57}{4}f_1)~
f_1^{9/2}~.
\label{generaln}
\ee
Note that there is no independent integration constant for $f_0$.
The general solutions 
define a $n$-dimensional hypersurface
in the space of $n+1$ coupling constants, which
is nothing but the critical surface.
We see from  (\ref{deltafn}) that
thanks to the exponential function $\exp [(2-2n)/3f_1]$
the deviations $\delta f_n$ from the
power series solutions vanish very fast as $f_1$
approaches zero,
implying that the power series solutions (\ref{power0})
and (\ref{powern})
are very attractive in the infrared limit.
This infrared attractiveness is interpreted as perturbative 
renormalizability by Polchinski \cite{polchinski,becchi}.
We adapt to his interpretation, and call these solutions
the perturbative  solutions \footnote{
Except for the first coefficients in the expansions 
(\ref{power0}) and (\ref{powern}),
we cannot compare them with those
in perturbation theory, because
the approximation employed here to solve the 
nonperturbative evolution equation (\ref{rged4}) is different from 
that in perturbation theory.
See, however, the discussion below.}. The exponential deviations 
defined in (\ref{linear0})
are therefore nonperturbative contributions, which  cannot be
computed usually  \cite{luscher1}.
So we call them
a nonperturbative ambiguity as in Ref.~\cite{luscher1}.

Before we proceed, we would like to argue that the
nonperturbative ambiguity we mentioned above
results from the fact that perturbation series diverge,
that is, it is the renormalon ambiguity \cite{renormalon}.
We have computed higher orders in
the power series expansion (\ref{generaln})  and found that they 
do not approximate the exact (numerical) result better. 
The one with the first four terms in (\ref{generaln})  
is the best approximation among lower orders.
From this fact, we believe that 
the power series solution (\ref{generaln}) does not converge,
and that it is an asymptotic series.
So, the power series (\ref{generaln})
reflects the property of perturbation series in the conventional
perturbation theory, as far as our numerical
analysis in lower orders suggests.
This is one of the reasons why we interpret the power series 
as the perturbative series.  This interpretation is also supported 
by the fact that not only the leading form of the nonperturbative
ambiguity,
the last exponential term in (\ref{generaln}),
agrees with 
that of the known renormalon ambiguity
\cite{renormalon}, but also the coefficient
of $1/f_1$,  $2/3$,  in the exponential function.
The power of $f_1$ in front of the exponential
function, that is $9/2$, 
differs slightly from the expected value $3$.
The origin is presumably the local potential approximation 
to the exact RG evolution equation.
Therefore, we believe that 
the last term in (\ref{generaln})
is the renormalon ambiguity.
As we will see, we can remove this ambiguity
by requiring the maximal  cutoff.

Off the critical surface (defined by (\ref{general0}) and (\ref{generaln})),
$f_0$ increases in the infrared limit, and approaches
infinity. We would like to solve the reduction equation (\ref{reduction})
in this limit. To this end, we consider the evolution
of $m^2=2 f_0 f_1$ (the mass squared in the broken phase)
and find that it also increases in the limit.
Therefore, in the lowest order, we just have to solve
\be
-\frac{3}{4} f_1^2 \frac{d f_n}{d f_1}
&\simeq & (-1)^n\frac{3}{4}f_1^{n+1}+(2-2n) f_n~,
\label{linear1}
\ee
and obtain for instance
\be
f_2
& \simeq &-\frac{4}{3}f_1-\frac{1}{2}f_1^2+\frac{32}{9}\exp(-\frac{8}{3f_1})
Ei(\frac{8}{3f_1})+\hat{K}_{2}\exp(-\frac{8}{3f_1})~.
\label{f2}
\ee
The solution becomes a power series in $f_1$, if the
integration constant 
$\hat{K}_{2}$ is  set equal zero\footnote{This can be seen
by using Soldner's theorem, 
$-\exp (1/x) Ei (-1/x)=1/(1+x/(1+x/(1+2x/(1+\cdots)$.}:
\be
f_2
&=&\frac{3}{8}f_1^3+\frac{9}{8}f_1^4+\cdots
\label{f2pt}
\ee
So the last term in (\ref{f2}) exhibits the nonperturbative ambiguity,
which we will compute by requiring the maximal cutoff later on.

To proceed, we would like  to study
the ultraviolet behavior of the theory, and consider the 
general form of the $\beta$ functions (\ref{bn})
as well as the explicit form for $\beta_0$ and $\beta_1$
(given in (\ref{a1})--(\ref{a4}))
along with the reduction equation (\ref{reduction}).
To solve (\ref{reduction}), we assume that $|f_n|~~(n \geq 2)$ 
increases as $f_1$ increases, while $f_0$ remains 
finite in the limit. It follows that under this assumption, 
the coupling constants have the leading
behavior
\be
f_0 &= &J_0^{(0)}+J_0^{(1)}/f_1+O(f_1^{-2})~,
\label{f01}\\
f_n &=& J_n^{(0)} f_1^{n}+O(f_1^{n-1})~,~(n \geq 2)~.
\label{fn1}
\ee
The coefficient $J_0^{(0)}$ cannot be 
determined, while
the others can be uniquely computed as functions of 
$J_0^{(0)}$. At $n=3$ for instance we find 
\be
J_0^{(1)} &=& 
\frac{-3+8 J_0^{(0)}-2 J_2^{(0)}}{3+8 (J_2^{(0)})^{2}-6 J_3^{(0)}}~,  
\label{j0}\\
J_2^{(0)} &=&\left\{ \begin{array}{c}
-1.2186\cdots\\0.3722\cdots \end{array}\right.~,~
J_3^{(0)} =\left\{ \begin{array}{c}
0.8382\cdots\\0.2329\cdots\end{array}\right.~~
\mbox{for}~~\left\{ \begin{array}{c}
\mbox{broken phase}\\\mbox{unbroken phase}\end{array}\right.~.
\label{j23}
\ee
The leading behavior given by (\ref{f01}) and (\ref{fn1})
is stable in the ultraviolet limit
at least for $n\leq 3$.
To show this, we construct
the general solutions that
surround the lading behavior (\ref{f01}) and  (\ref{fn1}) 
in the ultraviolet limit,
and denote the deviations from the leading behavior 
by  $\delta f_n$.
We first find 
\be
\frac{d\delta f_0}{d f_1} &\simeq & -r_0 \frac{\delta f_0}{f_1^2}~,
r_0\simeq \left\{ \begin{array}{c}
0.538\\ 1.95\end{array}\right.~~
\mbox{for}~~\left\{ \begin{array}{c}
\mbox{broken phase}\\ \mbox{unbroken phase}\end{array}\right.~.
\label{linear2}
\ee
This implies that  the leading order deviation $\delta f_0$ is just a shift 
of the undetermined constant $J_0^{(0)}$, which we denote
by $\delta  J_0^{(0)}$. For $\delta f_{2,3}$, we find
\be
\left(\begin{array}{c}
\frac{d\delta f_2}{d f_1} \\\frac{d\delta f_3}{d f_1}
\end{array}\right) &\simeq &
\left(\begin{array}{cc}
-4.98/f_1~, \  -2.95/f_1^2\\
-9.57~, \ -3.98/f_1
\end{array}\right)
\left(\begin{array}{c}
\delta f_2\\ \delta f_3
\end{array}\right) 
+\delta  J_0^{(0)}\left(\begin{array}{c}
4.02/(J_0^{(0)})^2\\ 
5.72 f_1/(J_0^{(0)})^2
\end{array}\right) 
\label{linear3}
\ee
for the broken phase. Similarly, we find that for the unbroken phase
\be
\left(\begin{array}{c}
\frac{d\delta f_2}{d f_1} \\\frac{d\delta f_3}{d f_1}
\end{array}\right) &\simeq &
\left(\begin{array}{cc}
0.302/f_1 ~,\  3.26/f_1^2\\
-1.03 ~,\ 1.30/f_1
\end{array}\right)
\left(\begin{array}{c}
\delta f_2\\ \delta f_3
\end{array}\right) 
+\delta  J_0^{(0)}\left(\begin{array}{c}
0.0697/(J_0^{(0)})^2\\ 
0.0491 f_1/(J_0^{(0)})^2
\end{array}\right) ~.
\label{linear4}
\ee
The corresponding solutions are found to be
\be
\delta f_2 &\simeq & \left\{ \begin{array}{l}
0.101 (\eta_a+\eta_b)\delta  J_0^{(0)}f_1/(J_0^{(0)})^{2}+
\eta_a f_1^{-10.3}+\eta_b f_1^{0.329}\\
\\  0.0185 (\eta_c+\eta_d)\delta  
J_0^{(0)}f_1/(J_0^{(0)})^{2}+f_1^{0.302}
[~\eta_c\cos(1.83\ln f_1)\\+
\eta_d\sin(1.83 \ln f_1)~]
\end{array}\right.~,
\label{deltaf2}\\
\delta f_3 &\simeq & \left\{\begin{array}{l}
1.19 (\eta_a+\eta_b)\delta  J_0^{(0)}f_1^2/(J_0^{(0)})^{2}+
1.80 [\eta_a f_1^{-9.30}+\eta_b f_1^{1.33}]
\\  \\
 -0.0231 (\eta_c+\eta_d)\delta  J_0^{(0)}f_1^2/(J_0^{(0)})^{2}
 -0.562f_1^{1.30}
[~\eta_c\sin(1.83\ln f_1)\\-
\eta_d\cos(1.83\ln f_1)~]
\end{array}\right.
\label{deltaf3}
\ee
for the broken and unbroken phases, respectively.
Therefore, $\lim_{f_1\to \infty}\delta f_2/f_2=
\lim_{f_1\to \infty}\delta f_3/f_3=0$, implying that the leading 
behavior (\ref{f01}) 
and (\ref{fn1}) is stable, and 
that the coupling constants
$f_2$ and $f_3$ (which have a negative
canonical dimension before they were made dimensionless)
diverge at the same scale $\Lambda_0$ at which  $f_1$ diverges.
This feature presumably continues to higher order truncations.
That is, all the coupling constants $f_n~~(n\geq 1)$ diverge
at the common scale.
But whether our method works or not does not depend
on it.

At this stage we would like to explain why we have 
shifted $\rho$ in the ansatz (\ref{ansatz}).
The reason is the following.  $f_0$ approaches
a finite,  undetermined constant $J_0^{(0)}$ in the ultraviolet
limit so that
the propagator effect $\Delta$ (given in (\ref{delta}) )
which appears in the $\beta$ functions $\beta_n$
is finite for any positive $f_1$ in the ultraviolet limit.
This makes the numerical analysis in the ultraviolet
limit easier, while without the shift in the unbroken case 
$\Delta=(1+m^2)^{-1}$ may easily become singular for some finite positive
$f_1$. Of course this complication is
not a serious hindrance for applying  our method.

According to Sect. II, we regard $f_0$ and $f_1$ as independent
parameters, while the other coupling 
constants $f_n ~~(n \geq 2)$ should be regarded as dependent.
We then require  that
for  given values of $f_0$ and $f_1$ at
$\Lambda << \Lambda_0$,
all the dependent coupling constants $f_n~~(n\geq 2)$ should 
be so fine tuned at $\Lambda$ that $f_n ~~(n \geq 1)$
diverge at the maximal value
of $\Lambda_0$.
This program cannot be solved analytically, and we relay 
on numerical analyses.
In Fig.~2 we plot the running time
$T_0=\ln(\Lambda_0/\Lambda)$ against $f_2(t=0)$ 
for $f_0(0)=1/4a=4\pi^2$ and $f_1(0)=0.1$ in the case of the truncation at $n=2$.
Comparing Fig.~2 with Fig.~1 of the nontrivial case, 
we observe a similarity, although $T_{\rm max}$ cannot be infinite
in the present case.
We see that $T_0$ is peaked at $f_2(0) \simeq 0.000528$.
To see the truncation dependence we calculate the fine-tuned value
of $f_2(0)$  as a function of
$f_1(0)$ for the truncations at $n=2$ and $3$.
The result is plotted in Fig.~3 for the case $f_0(0)=1/4a=4\pi^2$.
The solid and dotted  lines correspond
to  $n=2$ and $3$, respectively.
We observe that the result does not depend very much on the order of
truncation. 
The fine-tuned value
of $f_3(0)$  as a function of
$f_1(0)$ 
is plotted in Fig.~4 for  $f_0(0)=1/4a=4\pi^2$.

Now we come to the determination of
the coefficient $K_{2}$ in 
(\ref{generaln}), which exhibits
a nonperturbative correction of the renormalon 
type \cite{renormalon,luscher1}.
In Fig.~5 we plot
$T_0=\ln(\Lambda_0/\Lambda)$ against $K_2$, and in Fig.~6,
$T_0$ is plotted against $f_2(0)$ at $f_1(0)=0.1$.  From this result we obtain
\be
K_{2} &\simeq  7 \times 10^3~.
\label{k2hat}
\ee
This means a departure of about  $3~(0.1)$ \%  from the perturbative result
at $f_1=0.1~(0.07)$. Needless to say that  the
corresponding effect  in the standard model could be in principle measurable.

At last we would like to compare the maximal running
time $T_{\rm max}$ obtained in perturbation theory with
our nonperturbative value.
To this end, we use the perturbative evolution equation
\be
\frac{\partial f_1}{\partial t} &=& -3 f_1^2~,
\ee
and obtain
$T_{\rm max}=10/3\simeq 3.3$, which should be compared
with the nonperturbative result in Figs.~5 and 6.

These results obtained above show that our method works
in the four-dimensional case. It is certainly possible
to formulate our idea in terms of lattice theory.
Then one could improve the accuracy of the calculation
of $K_2$.

\subsection{$d=5,6$: Perturbatively unrenormalizable cases}
We now come to discuss unrenormalizable 
cases \footnote{There exist perturbatively unrenormalizable
theories which are nevertheless  nontrivial. 
See for instance Ref.~\cite{eguchi}.}:
\be
d=5,6  ~~\mbox{and}~~N=4~.
\ee
We emphasize that as far as renormalization is concerned,
and we are interested in the infrared regime, 
the scalar theory in five or higher dimensions may be seen as an 
oversimplified model of 
a low-energy effective theory of quantum gravity \cite{weinberg3}.

As in the case for $d=4$, we consider the derivative
$F=V'$, and follow basically the investigations of that case.
The calculations for the $d=6$ case
are the same as for the $d=5$ case, and
therefore, we consider only
the $d=5$ case below. So the details for the $d=6$ case
are suppressed below, but we give the final result 
for this case, too, at the end.
We first investigate
 the infrared behavior, and find that
the Gaussian point ($f_0=1/2~,~f_n=0 ~(n \geq 1)$)
is a fixed point. 
As in the previous case,
we construct explicitly the solution near the Gaussian fixed point
to investigate its stability. 
We find that the power series solutions like (\ref{power0})
and (\ref{powern}) do not exist. Instead,
the expansion of the form (which is similar 
to (\ref{sold3}) in the nontrivial case) 
\be
 f_0 &=&   \frac{1}{2}-\frac{3}{16} f_1+
\frac{171}{160} f_1^2+[~\frac{9}{2} \ln f_1 
-\frac{435}{128}+\frac{C_2}{12}~]~f_1^3
-[~\frac{54}{7} \ln f_1 
+\frac{554013}{62720}+\frac{C_2}{7}~]~f_1^4\nn\\
& &+[~\frac{3225933}{250880 }+ \frac{10441}{8960} C_2
+ \frac{281907}{4480}\ln f_1~]f_1^5+
O(f_1^6 \ln f_1)~,
\label{general05}\\
f_2 &=& f_1^{3}\left(\frac{15}{2}+[~54 \ln f_1 
+C_2~]~f_1+
[~81 \ln f_1 +\frac{3}{2}C_2-\frac{9315}{32}~]~f_1^2+O(f_1^3 \ln 
f_1)\right)~,
\label{generaln5}
\ee
for $d=5$ at the truncation $n=2$
exhibit the general solutions,
where $C_2$ is an integration constant.
As in the case for $d=4$, there is no independent integration constant for 
$f_0$, and 
the general solutions
define the $n$-dimensional critical surface
in the space of $n+1$ coupling constants.
As $K_2$ in (\ref{generaln}), we will calculate $C_2$
 by requiring the maximal cutoff later.
The ambiguity expressed by  $C_2$ 
in (\ref{generaln5}) 
is indeed suppressed 
by $f_1^{4}$ in the infrared limit, but
we also see that in contrast to 
(\ref{general0}) and 
(\ref{generaln}) in $d=4$ dimensions,
there are no strong infrared attractive
functions like the perturbative solutions 
(\ref{power0}) and  (\ref{powern}).
Therefore, following  Polchinski \cite{polchinski,becchi}, the theory is
perturbatively unrenormalizable, in accord with the result
in perturbation theory.

Whether  power series solutions like (\ref{power0}) and  (\ref{powern}) exist
depends crucially on the existence of 
a dimensionless coupling constant.
Namely, if there is no dimensionless coupling constant, 
there are no couplings for which the linear term
in the $\beta$ function,
$(2+2n -nd) f_n$ in the case of (\ref{bn}), vanishes.
Then the reduction equation (\ref{reduction}) 
 \footnote{Application of reduction of
coupling constants to quantum gravity has 
been considered in Ref.~\cite{atance}. See also \cite{atance2}.}
becomes
\be
d_x x \frac{d y}{ d x} &= &d_y y
\label{dxy}
\ee
in the vicinity of the Gaussian fixed point,
where $x$ and $y$ denote generic coupling constants, and $d_x$ 
and$d_y$ are their canonical dimensions before the dimensional
rescaling (\ref{scaling1}). The solution of (\ref{dxy}) is simply 
$y=\tilde{C }x^{d_y/d_x}$,
where $\tilde{C }$ is an integration constant, but one can convince
oneself that logarithmic terms such as  $x^{d_y/d_x}\ln x$ are needed
to construct a solution of the full problem. This is the  origin of 
the logarithmic terms in (\ref{general05}) and (\ref{generaln5}),
and also in (\ref{sold3}). The difference between $C_2$ in 
(\ref{generaln5})
and $C$ in (\ref{sold3}) is that $C$ is determined by
nontriviality, while $C_2$ will be determined by the maximal cutoff.

Off the critical surface,
the mass squared  $m^2=2 f_0 f_1$ in the broken phase
increases in the infrared limit, and approaches
infinity. 
In the lowest order, we therefore have to solve
\be
-f_1\frac{d f_n}{d f_1}
&\simeq & (-1)^n\frac{3}{4}(1+3^n)f_1^{n+1}+(2-3n) f_n~,
\label{linear15}
\ee
and obtain 
\be
f_n
&\simeq &\frac{3}{4(3-2n)}(-f_1)^{n+1}+\hat{C}_n f_1^{3n-2}~
~,~(n \geq 2)~.
\label{f25}
\ee
So the last term in (\ref{f25}) exhibits the nonperturbative ambiguity,
which could be removed by  our method.

The ultraviolet behavior in  the case for $d=5$ is basically the same
as in the case for $d=4$. Indeed, except for (\ref{j0}) and (\ref{linear2}),
which now become
\be
J_0^{(1)} &=& 
\frac{-3+12J_0^{(0)}-2 J_2^{(0)}}{3+8 (J_2^{(0)})^{2}-6 J_3^{(0)}}~, 
\\ 
\label{j05}
\frac{d\delta f_0}{d f_1} &\simeq & -r_0 \frac{\delta f_0}{f_1^2}~,
r_0=\left\{ \begin{array}{c}
0.8064\cdots\\2.9208\cdots\end{array}\right.~~
\mbox{for}~~\left\{ \begin{array}{c}
\mbox{broken phase}\\ \mbox{unbroken phase}\end{array}\right.~,
\label{linear25}
\ee
all the results, (\ref{j23}) and  (\ref{linear3})--(\ref{deltaf3}),
remain valid. So concerning this part of our discussions,
we have nothing to add to what we 
have found in the case for $d=4$.

Only $f_0$ has a canonical 
dimension $\geq 0$ before the rescaling (\ref{scaling1}), and the 
coupling constant that has the largest canonical dimension 
among the 
coupling constants with a negative canonical dimension is $f_1$.
Therefore, according to Sect. II,  we regard $f_0$ and $f_1$ as independent
parameters, while the other coupling 
constants $f_n ~~(n \geq 2)$ should be regarded as dependent.
Then we require  that
for given values of $f_0$ and $f_1$ at
$\Lambda << \Lambda_0$,
all the dependent coupling constants $f_n~~(n\geq 2)$ should 
be so fine tuned at $\Lambda$ that $f_n ~~(n \geq 1)$
diverge at the maximal value
of $\Lambda_0$.
We perform similar numerical investigations as  in the case for $d=4$.
In Figs.~7 and 8 we plot $T_0=\ln(\Lambda_0/\Lambda)$ against $f_2(0)$ 
for $f_0(0)=10.0$ and $f_1(0)=0.1$ in the case of the truncation at $n=2$
for $d=5$ and $d=6$, respectively.
We see that $T_0$ is peaked at $f_2(0)\simeq 0.0015$ for $d=5$
and $f_2(0)\simeq 0.002 $ for $d=6$, respectively.
As we can see from Figs.~7 and 8, the maximal 
value $T_{\rm max}$ is relatively low
compared with the four-dimensional case (Figs.~2, 5 and 6).
To increase $T_{\rm max}$, we of course have to decrease
 $f_1(0)$.
The the fine-tuned value
of $f_2(t=0)$ is plotted as a function of
$f_1(0)$ for the truncation at $n=2$
with $f_0(0)=1/4 a$ in Figs.~9 and 10
for $d=5$ and $d=6$, respectively.

In Fig.~11 we plot the running time
$T_0=\ln(\Lambda_0/\Lambda)$ 
on the critical surface against $C_2$ for $d=5$, where the
nonperturbative ambiguity $C_2$ is defined in (\ref{generaln5}).
As for $d=6$ the general solutions corresponding
to (\ref{general05}) and (\ref{generaln5}) become
\be
 f_0 &=&   \frac{3}{8}-\frac{3}{32} f_1+
[~\frac{153}{1024}+\frac{3}{64}C_2 -\frac{45}{256}\ln f_1~]f_1^2
+[~-\frac{9}{1204} \ln f_1 
+\frac{2745}{4096}+\frac{3}{1280}C_2~]~f_1^3\nn\\
& &+[~-(\frac{1089}{16384} +\frac{135}{512}C_2 )\ln f_1 
+\frac{2025}{4096}(\ln f_1)^2 
-\frac{10005}{8192}+\frac{363}{20480}C_2
+\frac{9}{256}C_2^2~]~f_1^4\nn\\
& &+O(f_1^5 (\ln f_1)^2)~,\\
\label{general06}
f_2 &=& f_1^{3}\left( C_2-\frac{15}{4}\ln f_1+[~-\frac{135}{16}\ln f_1 
+\frac{693}{32}+\frac{9}{4}C_2~]~f_1
+O(f_1^2 (\ln f_1)^2) \right)~.
\label{generaln6}
\ee
Fig.~12 shows the running time $T_0$ on the critical surface
against $C_2$ for $d=6$, from which we
obtain
\be
C_2 &\simeq  1.1 \times 10^2~(-7.6)~~\mbox{for} ~~d=5 ~(6)~.
\label{k22}
\ee
In Fig.~13 and 14 we plot
the running time $T_0$ on the critical surface
as a function of the $f_2(0)$, where $f_1(0)$ is fixed at
$0.07$ and $0.15$ for $d=5$ and $6$, respectively.

Using the perturbative evolution equation
\be
\frac{\partial f_1}{\partial t} &=& (4-d) f_1-3 f_1^2~,
\ee
we would obtain
$T_{\rm max}\simeq 1.75$ and $0.85$ for
$d=5$ and $6$, which should be compared
with the nonperturbative results in Fig.~13 and 14, respectively.
In more realistic situations, the running time should be
understood as $T_0=\ln (\Lambda_0 R)$, where $R$ is
a length scale of the compactification of extra dimensions.
Our calculations show that $T_0$ depends
on the values of the coupling constants of 
higher-dimensional operators at $1/R$.
Therefore, it is possible to find phenomenological 
constraints-triviality constraints-on these coupling constants.

At last we 
would like to emphasize that our method
can be applied to more physically
interesting theories such as quantum gravity and
Yang-Mills theories in higher dimensions, regardless of whether
they have a nontrivial fixed point
\footnote{See for instance Refs. \cite{peskin} -\cite{egawa,souma}
 for nonperturbative
investigations on the existence 
of a nontrivial fixed point in theses theories.
Formulations of the Wilsonian 
RG in gauge theories and related studies have been  made
by many authors in recent years \cite{becchi,bonini}.
Phenomenological applications of the Wilsonian 
RG to higher-dimensional Yang-Mills theories
have been considered  in Res. \cite{kobayashi}.}.
If there exists no nontrivial fixed point
in the low-energy effective theory
of quantum gravity, the maximal value
of the intrinsic cutoff might depend on
the background geometry. Then it would be
interesting to see whether our criterion
will select a specific background geometry.

\section{Conclusion}
Since  Kaluza and Klein  \cite{kaluza} showed that the fundamental 
 forces can be unified  by introducing
extra dimensions, their idea has attracted attention for
many decades.
 Recently, there have been again growing interests in
 field theories in extra  dimensions \cite{arkani1}--\cite{dienes1}.
 In contrast to 
 previously suggested
Kaluza-Klein theories in which the size of 
extra dimensions was  of the order
of the Planck length or 
the inverse of the unification scale,
 the length scale of 
the extra dimensions in recent theories 
can be so large that they could  be experimentally observed.
Quantum corrections may also be observable.
But field theories in more than four dimensions are 
usually unrenormalizable. 
However, how to control them in unrenormalizable theories
is less known. 

We have addressed this problem in this paper, and
applied the Wilsonian RG to unrenormalizable theories.
We have assumed the existence of the maximal ultraviolet cutoff
in a cutoff theory, and required that the theory should be so 
adjusted that one arrives at the maximal cutoff.
We have applied
our method to the scalar theory in four, five and six dimensions.
The peaks that we have
seen in many figures in Sec. III
mean that  a particular RG flow can be  selected from
our requirement.
Based on this finding, we  would like to
conclude that these unrenormalizable
theories can obtain a predictive power in the infrared regime.
Although we have used
only the Wegner-Houghton  equation (\ref{wh}) in the 
derivative expansion approximation in   lower orders,
we believe the existence of such peaks in the exact result.
In other words, we believe that
unrenormalizable theories can possess a 
renormalization-group theoretical structure,
like perturbative renormalizability in a perturbatively renormalizable
theory, that enables us to make  unique predictions
in the infrared regime in terms of a finite number of independent
parameters.
 
The RG flow,  selected from our requirement,
runs for the maximal  time, and 
in the low-energy regime
this particular RG flow could be the best approximation 
to a renormalized trajectory
of the fundamental theory, for which the cutoff theory is the low-energy
effective theory.
Clearly, this idea can be tested in many different ways and 
field theory models.
Further results will be published elsewhere.

\acknowledgments

This paper may be seen as    an extension of 
reduction of coupling constants \cite{reduction,perry} to 
unrenormalizable  theories. In fact many techniques 
used and developed there have been applied in Sec. III.
One of us (J.K) would like to thank 
W. Zimmermann for instructive suggestions and discussions,
and a careful reading of the manuscript.
We would like to thank K-I. Aoki and H. Terao for useful discussions,
and also Y. Ikuta for numerical analyses.

This work is supported by the Grants-in-Aid
for Scientific Research  from 
 the Japan Society for the Promotion of Science (JSPS) (No. 11640266).
 
\appendix
\section{$\beta$ functions}
 
 We give here the $\beta$ functions of the coupling constants $f_n$
 for $N=4$ and $n \leq 3$, which are defined in (\ref{ansatz}).
 The $\beta$ functions are obtained by inserting the 
 expansion (\ref{ansatz}) into the evolution equation for $F$ in
 (\ref{rged4}):
 \be
\beta_0 &=& -\frac{3}{4}+(d-2)f_0-[~\frac{3}{4}
+\frac{f_0 f_2}{f_1}~]\Delta~,
\label{a1}\\
\beta_1 &=& (4-d)f_1-\frac{3}{4}f_1^2+
[~f_2-\frac{2 f_0 f_2^2}{f_1}+3 f_0 f_3~]\Delta-
[~\frac{9 f_1^2}{4}+6 f_0 f_1 f_2+4 f_0^2 f_2^2~]\Delta^2~,\\
\beta_2 &=& (6-2d)f_2+\frac{3}{4}f_1^3-\frac{9}{4}f_1 f_2+
[~3 f_3-\frac{3f_0 f_2 f_3}{f_1}+6 f_0 f_4~]\Delta \nn\\
& &-[~\frac{15}{2} f_1 f_2+10 f_0 f_2^2-9f_0 f_1 f_3+12 f_0^2 f_2 f_3~]
\Delta^2+ [~ \frac{27}{4} f_1^3-\frac{15}{4} f_1 f_2
+\frac{39}{2}f_0 f_1^2 f_2 \nn  \\
& &-5 f_0 f_2^2+26 f_0^2 f_1 f_2^2+
16 f_0^3 f_2^3-\frac{9}{2} f_0 f_1 f_3-9 f_0^2 f_1^2 f_3-
6 f_0^2 f_2 f_3-12 f_0^3 f_1 f_2 f_3~]\Delta^3~,\\
\beta_3 &=& (8-3d) f_3-\frac{3}{4}f_1^4-\frac{3}{2}f_2^2-3f_1 f_3 +3f_1^2 f_2
+[~10 f_0 f_5+6 f_4-\frac{4 f_0 f_2 f_4}{f_1}~]\Delta\nn\\
& &-[~\frac{63}{4}f_1 f_3+21 f_0 f_2 f_3+18 f_0 f_1 f_4
-24 f_0^2 f_2  f_4~]\Delta^2\nn\\
&& +[~\frac{45}{2}f_1^2 f_2-\frac{25}{2}f_2^2+35 f_0 f_1 f_2^2
+40 f_0^2  f_2^3+27 f_0 f_1^2 f_3\nn\\
& &-30 f_0 f_2 f_3+12 f_0^2 f_1 f_2 f_3+
48 f_0^3 f_2^2 f_3-18 f_0^2 f_3^2-36 f_0^3 f_1 f_3^2~]\Delta^3\nn\\
& &+[~-\frac{81}{4}f_1^4+\frac{45}{2}f_1^2 f_2-63 f_0 f_1^3 f_2+
60 f_0 f_1 f_2^2-96f_0^2 f_1^2 f_2^2
+40f_0 ^2 f_2^3-112 f_0 ^3 f_1 f_2^3\nn\\
& &-64 f_0 ^4  f_2^4-\frac{21}{4}f_1f_3+6 f_0 f_1^2 f_3+33 f_0^2 f_1^3 f_3-
7 f_0 f_2 f_3+44 f_0 ^2 f_1 f_2 f_3\nn\\
& &+116 f_0^3  f_1^2 f_2 f_3+48 f_0^3 f_2^2 f_3
+96 f_0^4 f_1 f_2^2 f_3-6 f_0 f_1 f_4-24 f_0 ^2 f_1^2 f_4-
24 f_0^3 f_1^3  f_4\nn\\
& &-8 f_0^2  f_2 f_4-32 f_0^3f_1 f_2 f_4-32 f_0^4 f_1^2 f_2 
f_4~]\Delta^4~,
\label{a4}
\ee
where $\Delta$ is defined in (\ref{delta}).

\newpage

\begin{figure}
\epsfxsize= 12 cm
\centerline{\epsffile{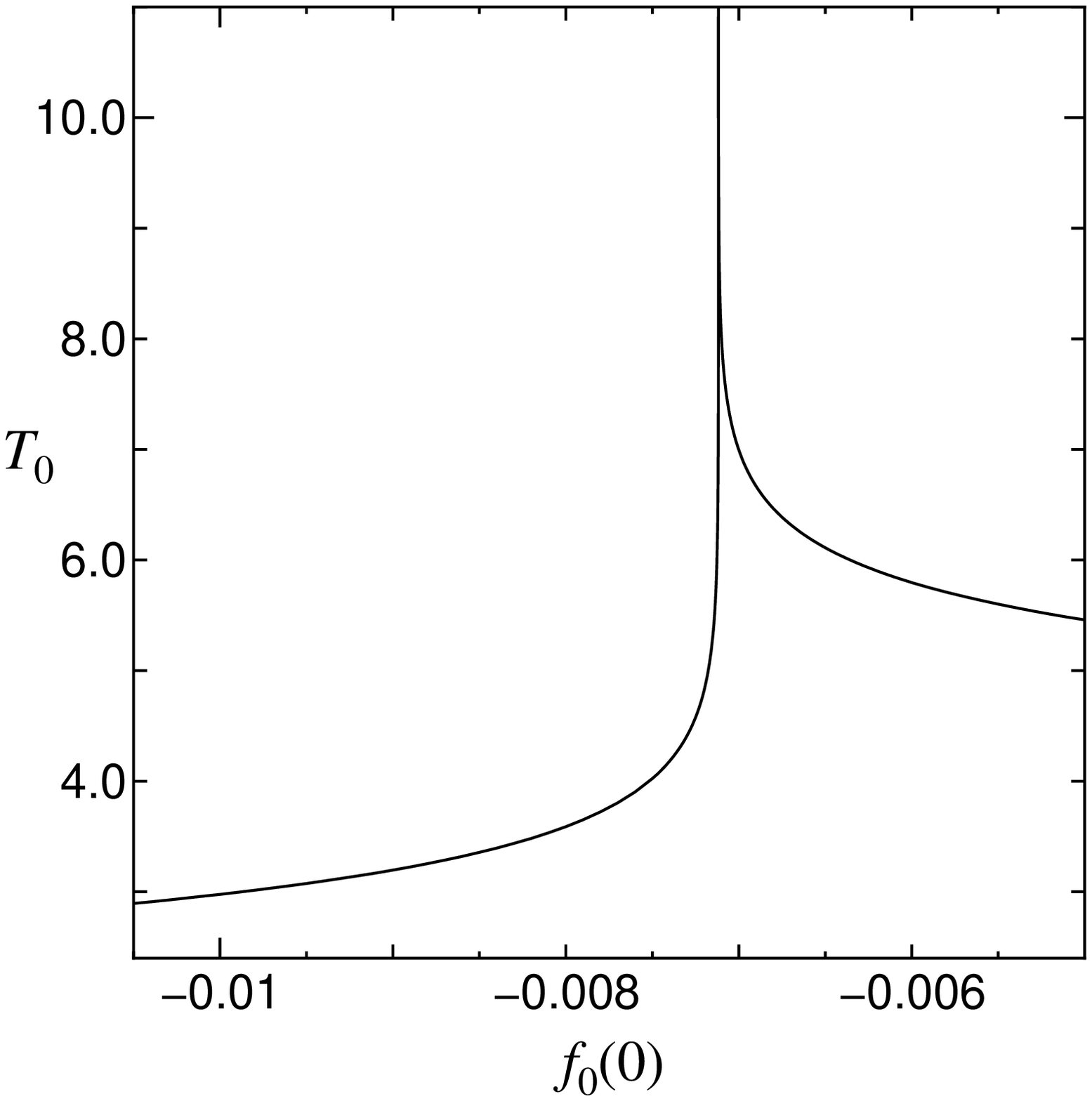}}
\caption{Fine-tuning of $f_0(0)$ for a given value
$f_1(0)=0.01$ in the
nontrivial case $d=3$ 
(with infrared and ultraviolet interchanged).
The ``running time'' $T_0=
\ln \Lambda_0/\Lambda$ becomes infinite at $f_0(0) \simeq -0.00718$.}
\label{fig:1}
\end{figure}

\begin{figure}
\epsfxsize= 12 cm
\centerline{\epsffile{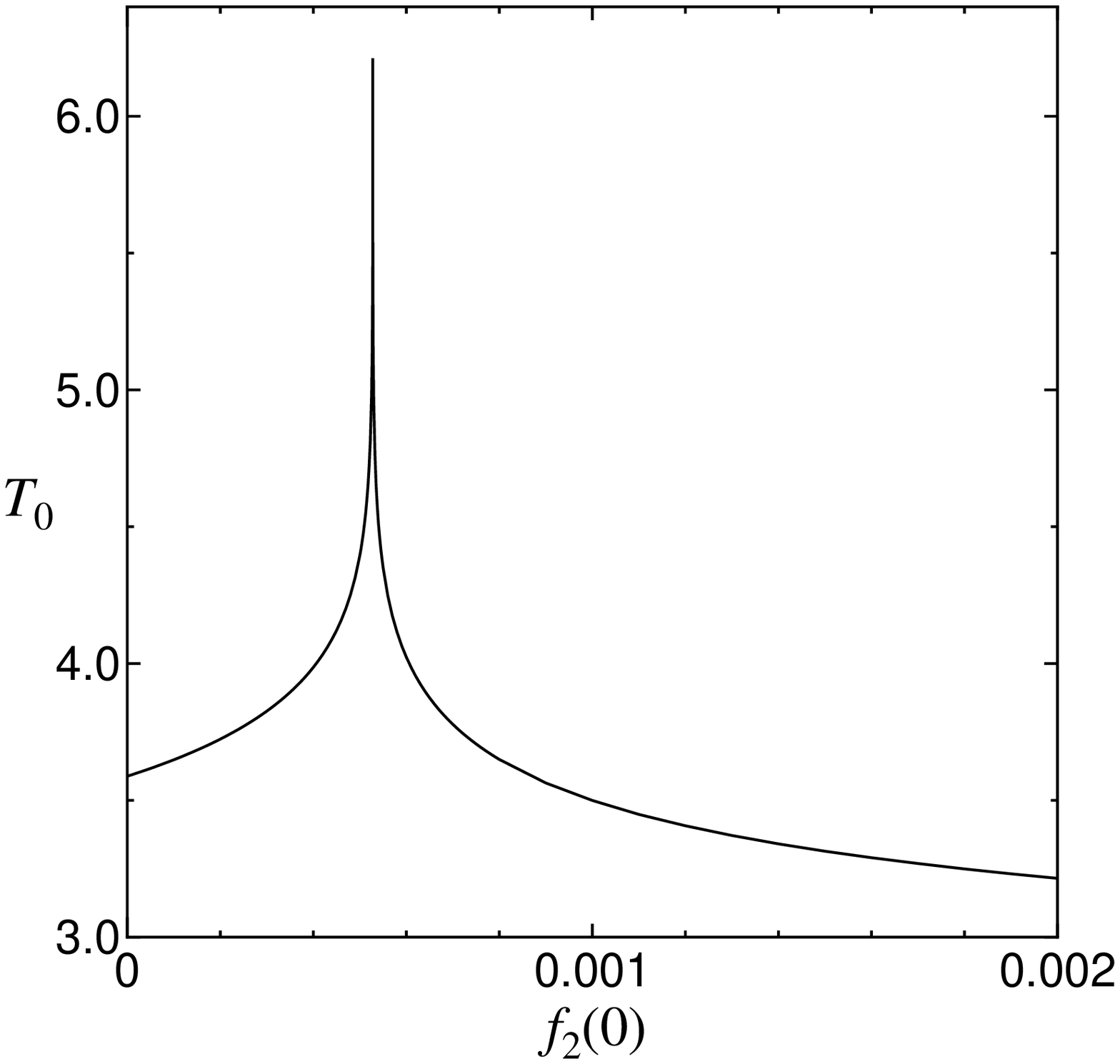}}
\caption{Fine-tuning of $f_2(0)$ for  given values
$f_1(0)=0.1$ and $f_0(0)=4\pi^2$
in the trivial, but perturbatively 
renormalizable case ($d=4$.)
The  running time $T_0$ becomes maximal at $f_2(0) \simeq 0.000528$.
$f_1$ and $f_2$ are the coupling constants
for $\phi^4$ and $\phi^6$, respectively
(see (\ref{ansatz})).  }
\label{fig:2}
\end{figure}

\begin{figure}
\epsfxsize= 12 cm
\centerline{\epsffile{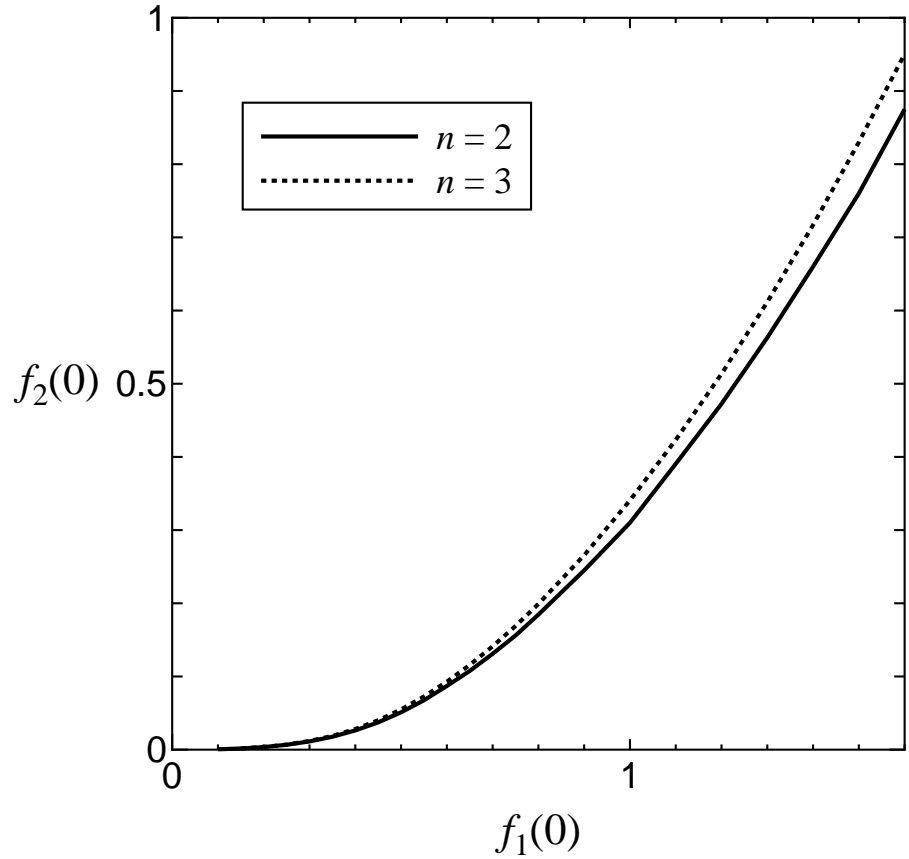}}
\caption{Fine-tuned value of $f_2(0)$ as a function of
$f_1(0)$  in $d=4$ ($f_0(0)=4\pi^2$).
The solid and dotted  lines correspond
to  the 
truncations at $n=2$ and $3$, respectively.}
\label{fig:3}
\end{figure}

\begin{figure}
\epsfxsize= 12 cm
\centerline{\epsffile{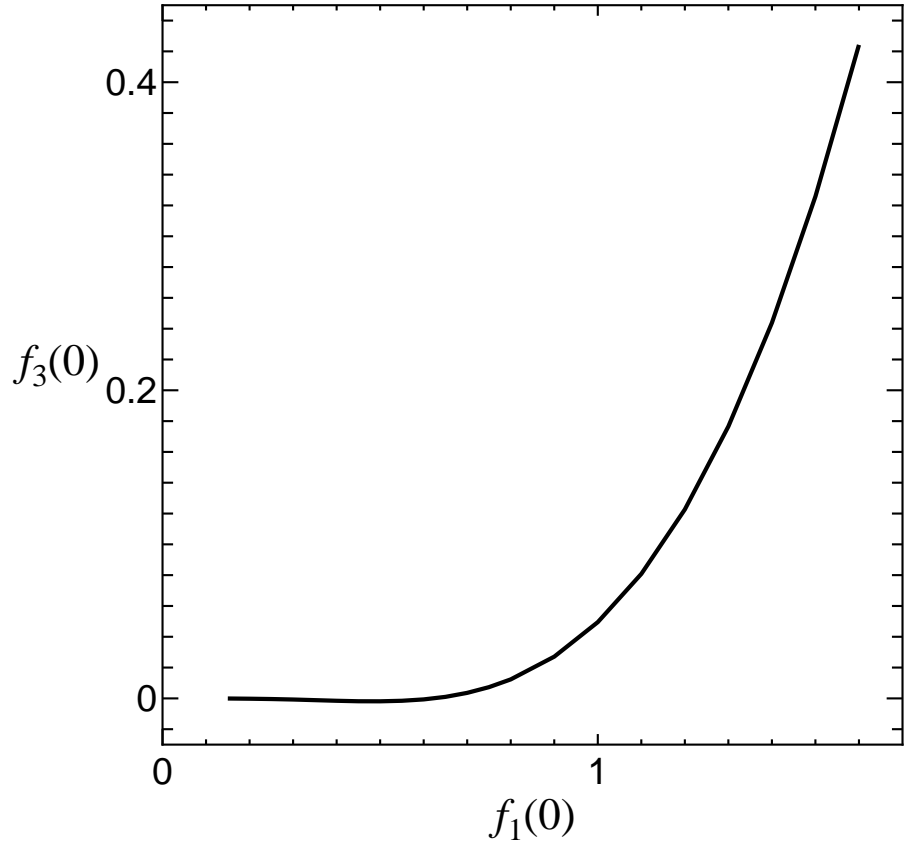}}
\caption{Fine-tuned value of $f_3(0)$ as a function of
$f_1(0)$  in $d=4$ ($f_0(0)=4\pi^2$).}
\label{fig:4}
\end{figure}

\begin{figure}
\epsfxsize= 12 cm
\centerline{\epsffile{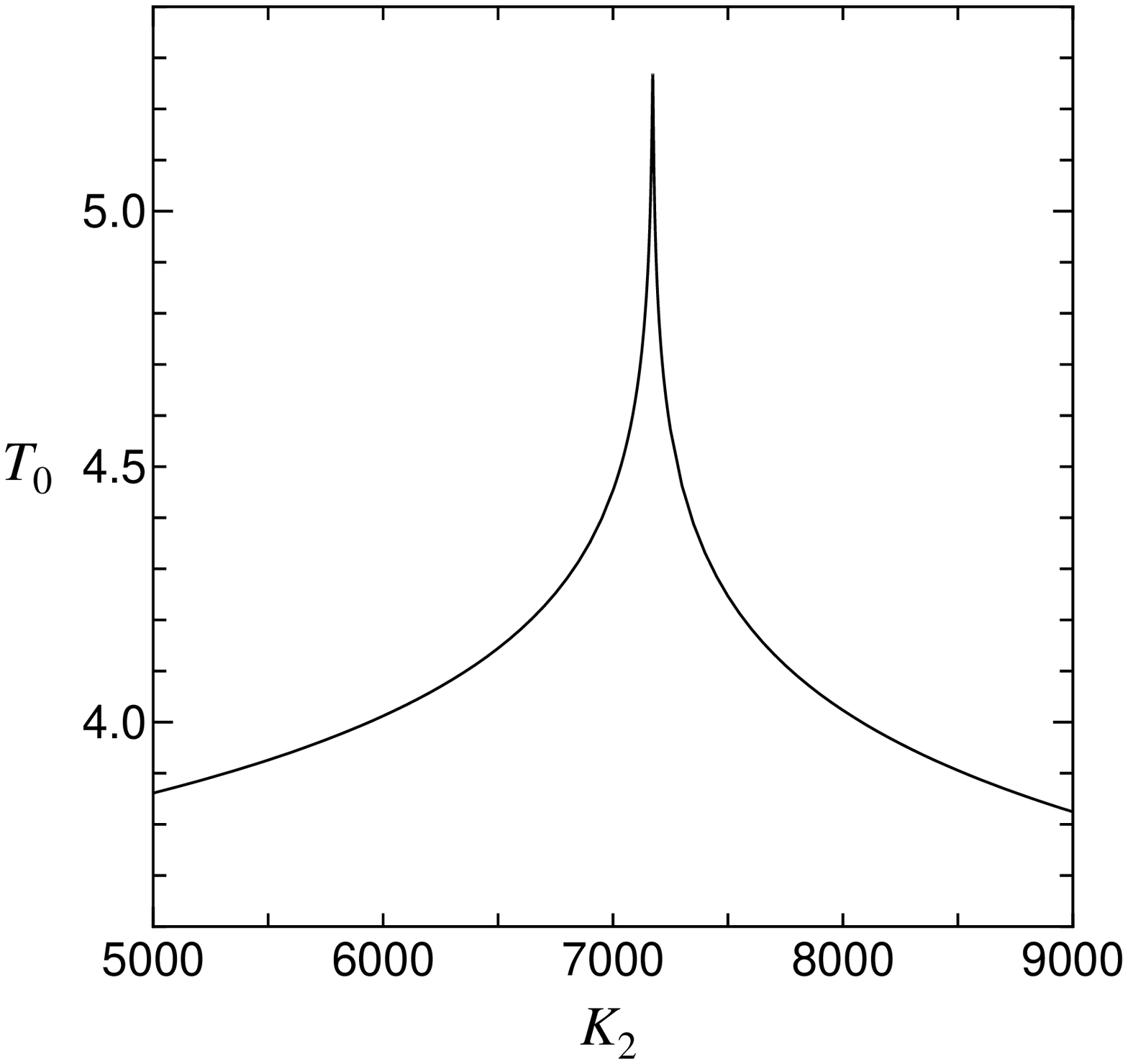}}
\caption{Determination of the nonperturbative coefficient $K_2$
(given in (\ref{generaln}))  in $d=4$.}
\label{fig:5}
\end{figure}

\begin{figure}
\epsfxsize= 12 cm
\centerline{\epsffile{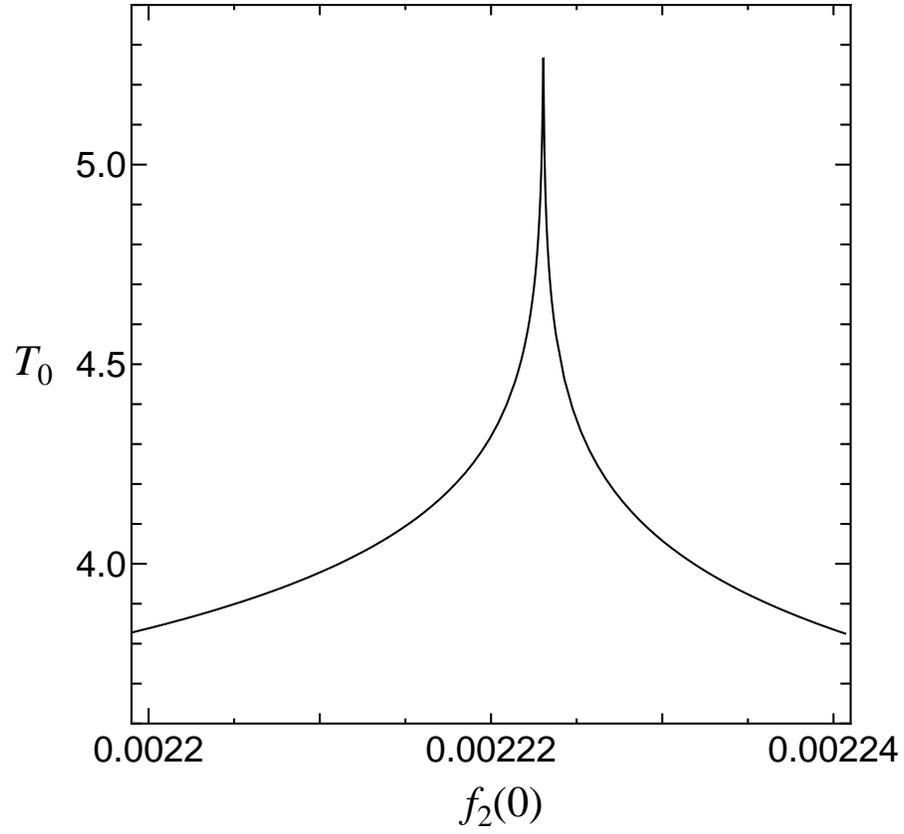}}
\caption{Removing the nonperturbative
ambiguity of $f_2$ on the critical surface at $f_1=0.1$ in $d=4$.}
\label{fig:6}
\end{figure}

\begin{figure}
\epsfxsize= 12 cm
\centerline{\epsffile{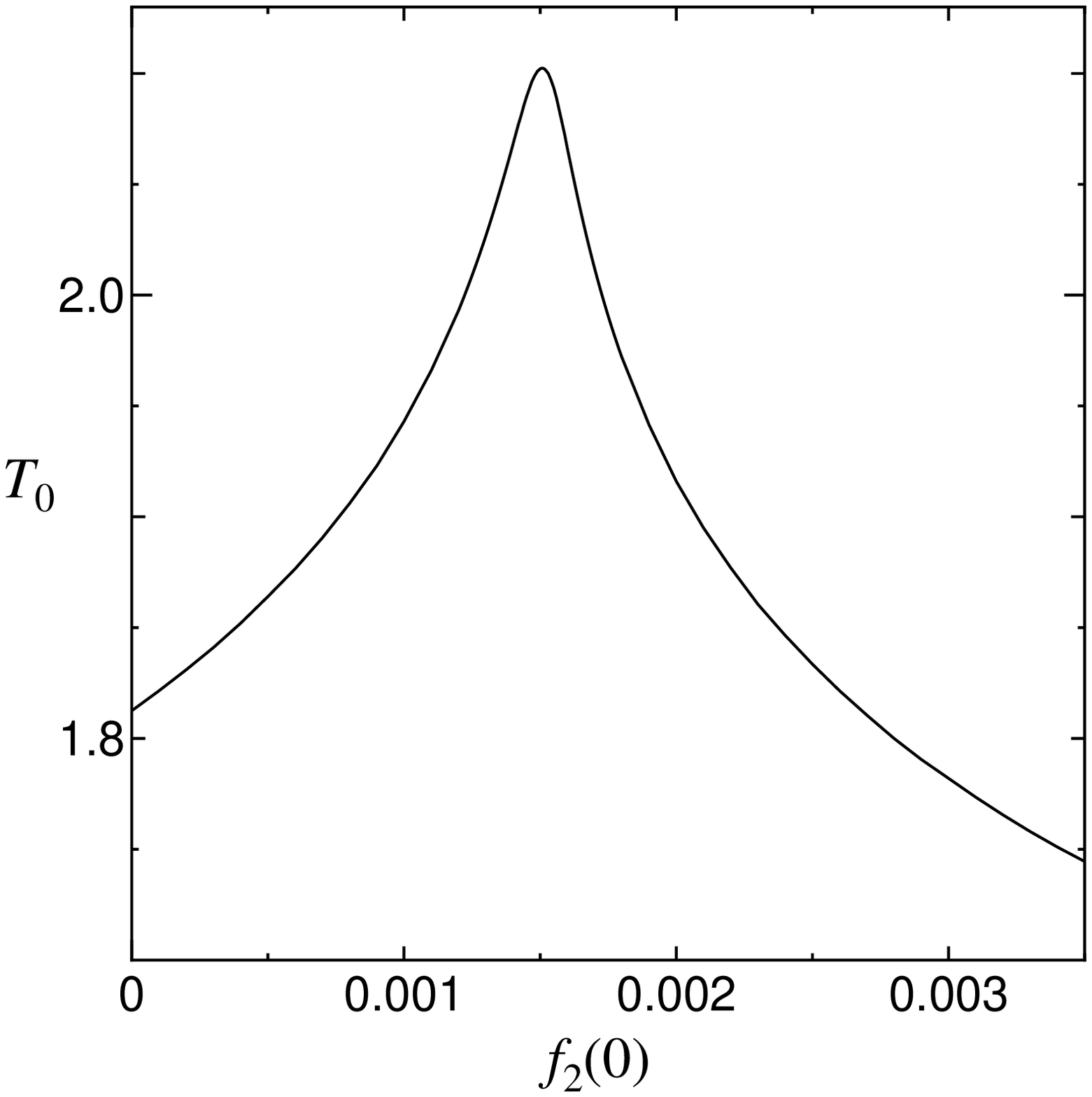}}
\caption{Fine-tuning of $f_2(0)$ for 
given values
$f_1(0)=0.1$ and $f_0(0)=10.0$
in the trivial,  perturbatively 
unrenormalizable case ($d=5$).
The  running time $T_0$ becomes maximal at $f_2(0) \simeq 0.0015$.
$f_1$ and $f_2$ are the coupling constants
for $\phi^4$ and $\phi^6$, respectively
(see (\ref{ansatz})).}
\label{fig:7}
\end{figure}

\begin{figure}
\epsfxsize= 12 cm
\centerline{\epsffile{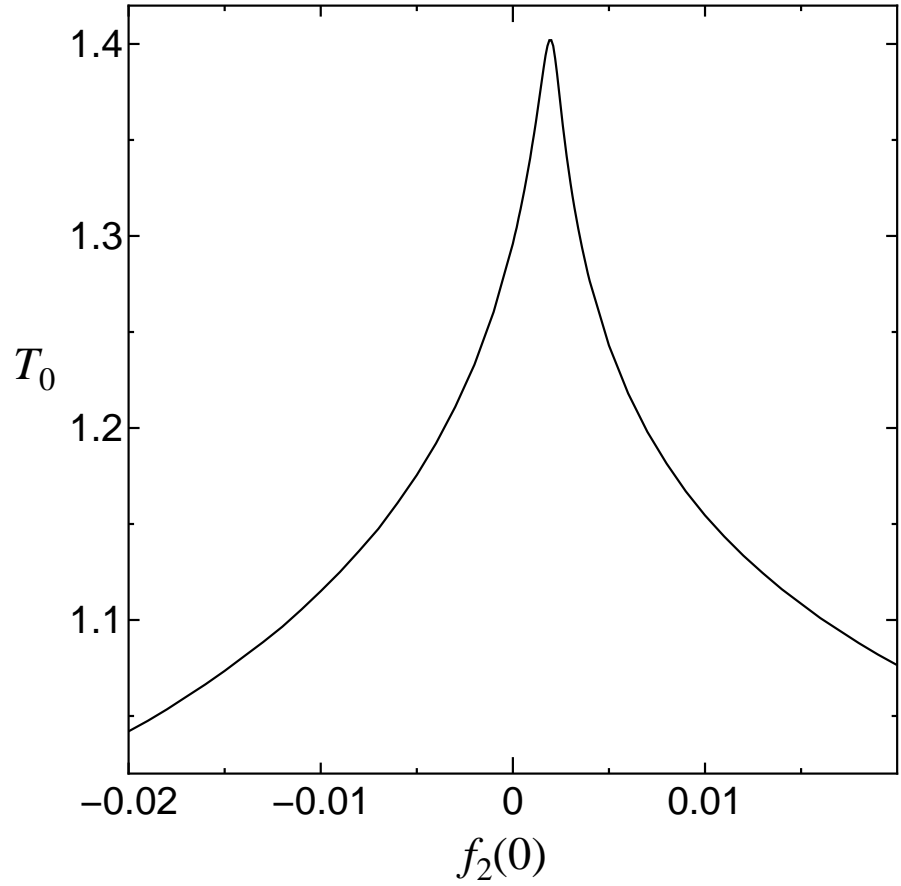}}
\caption{The same as
Fig.~6 for $d=6$.
The  running time $T_0$ becomes maximal at $f_2(0) \simeq 0.002$.}
\label{fig:8}
\end{figure}

\begin{figure}
\epsfxsize= 12 cm
\centerline{\epsffile{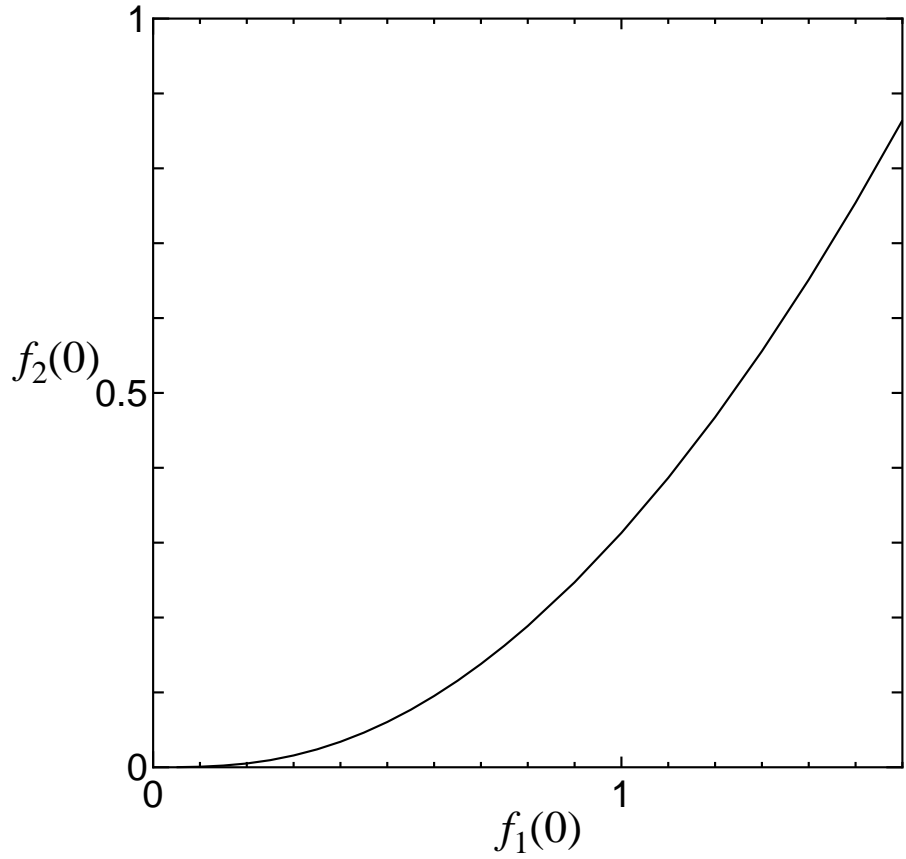}}
\caption{Fine-tuned value of $f_2(0)$ as a function of
$f_1(0)$ in $d=5$ ($f_0(0)=1/4 a$).}
\label{fig:9}
\end{figure}

\begin{figure}
\epsfxsize= 12 cm
\centerline{\epsffile{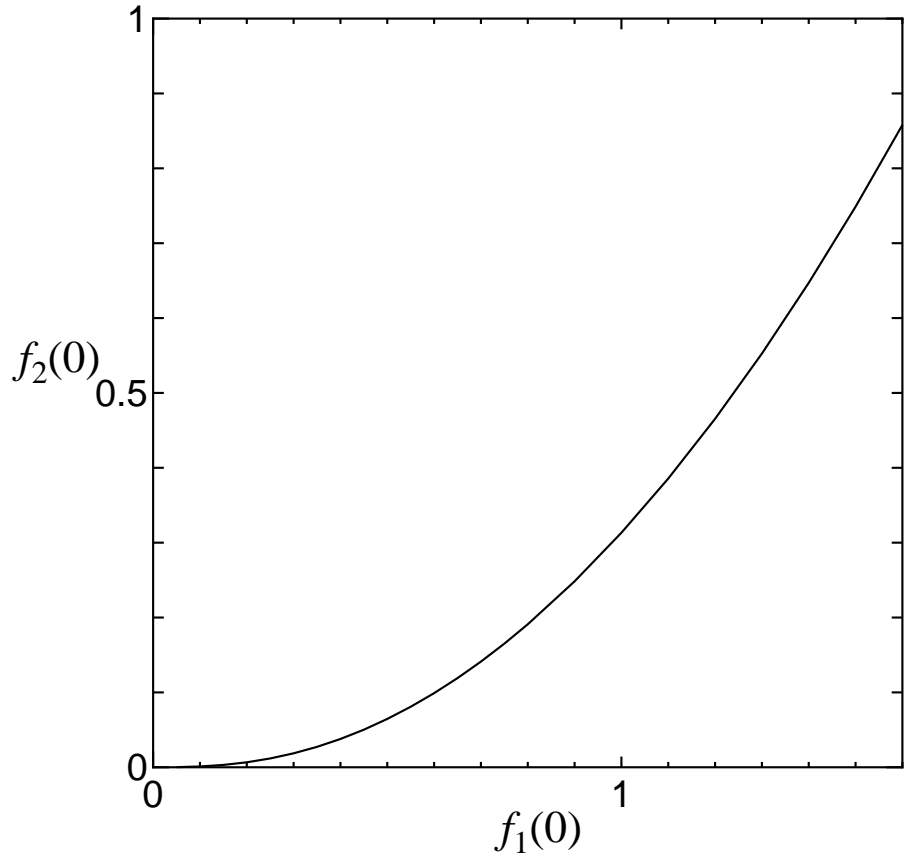}}
\caption{The same as Fig.~9  in $d=6$.}
\label{fig:10}
\end{figure}

\begin{figure}
\epsfxsize= 12 cm
\centerline{\epsffile{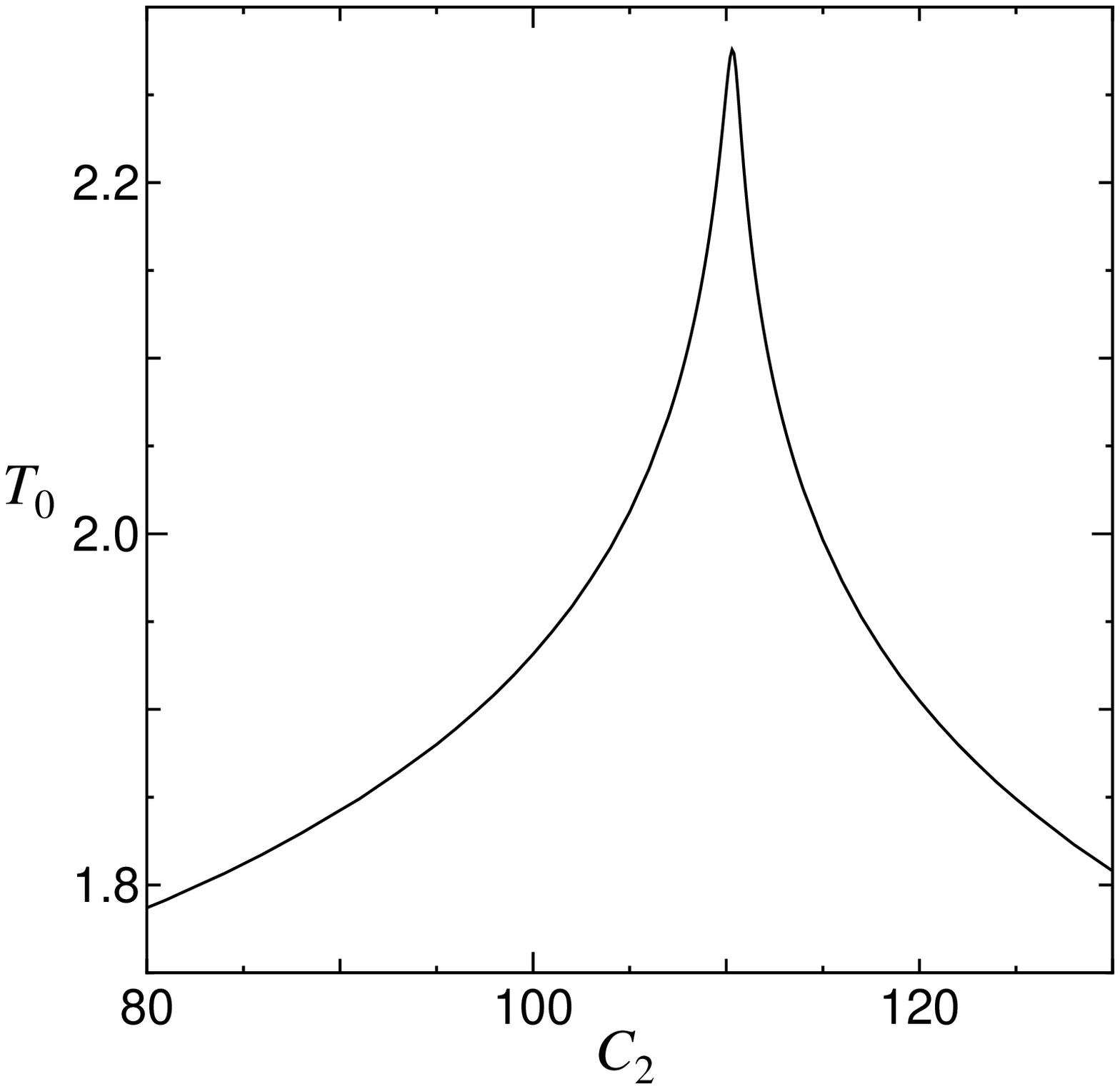}}
\caption{Determination of the nonperturbative coefficient $C_2$
(given in (\ref{generaln5}) ) in $d=5$.}
\label{fig:11}
\end{figure}

\begin{figure}
\epsfxsize= 12 cm
\centerline{\epsffile{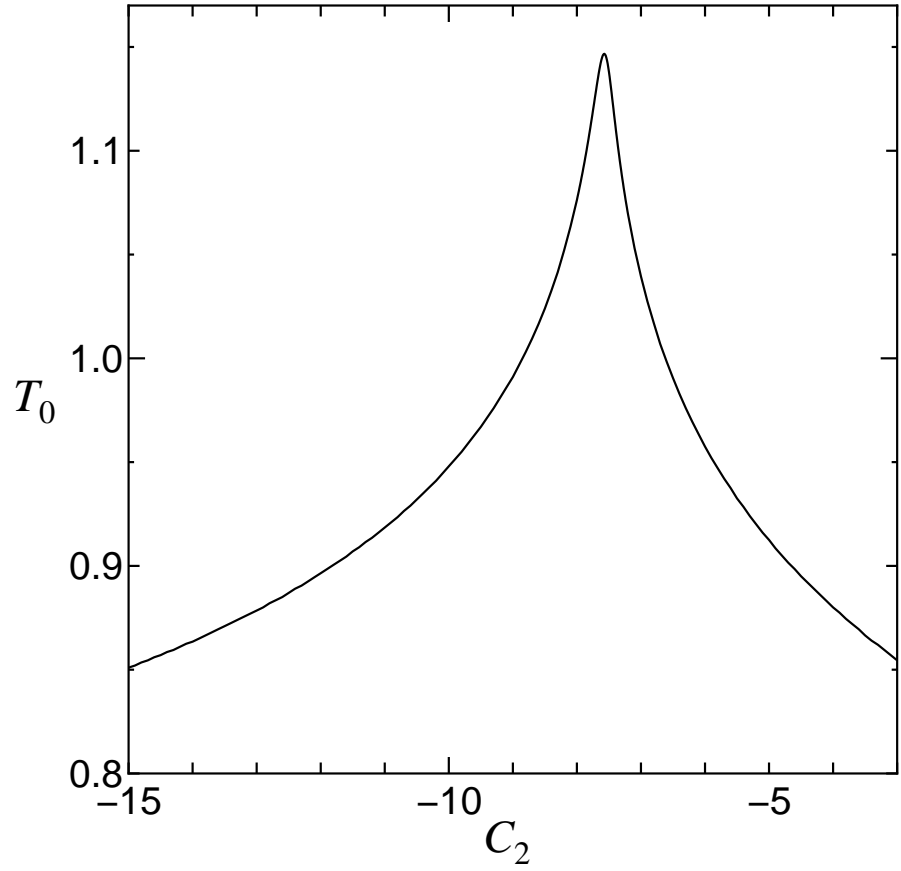}}
\caption{The same as Fig.~11 for $d=6$, where $C_2$ is given 
in (\ref{generaln6}).}
\label{fig:12}
\end{figure}

\begin{figure}
\epsfxsize= 12 cm
\centerline{\epsffile{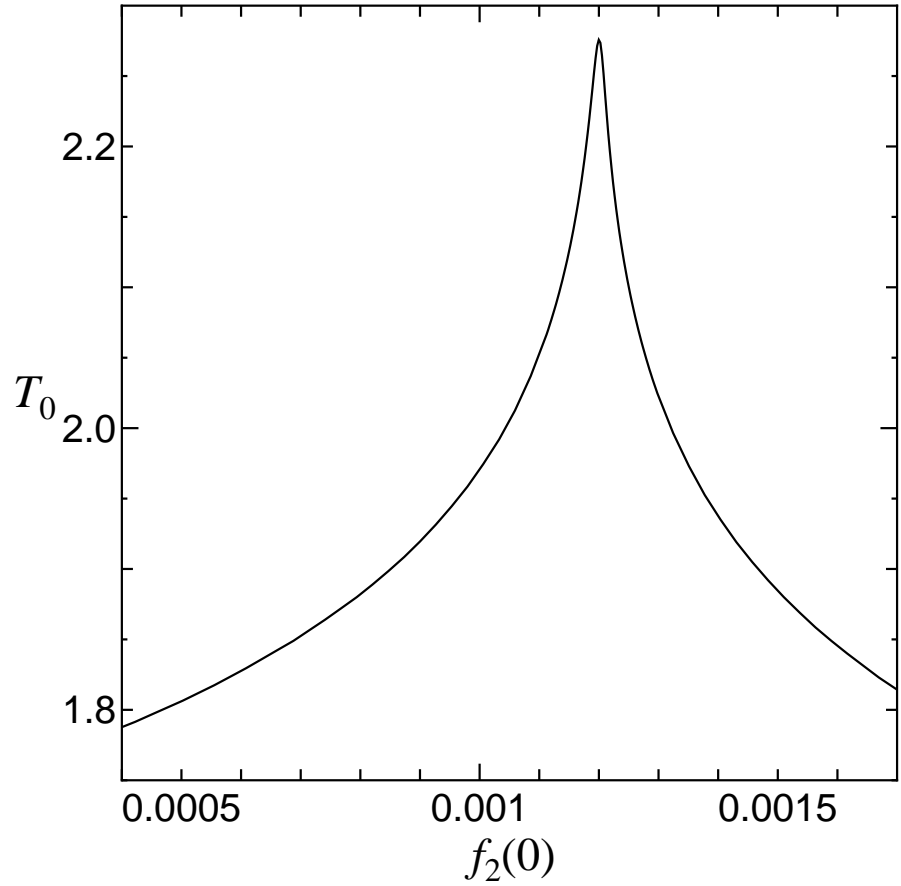}}
\caption{Removing the nonperturbative
ambiguity of $f_2$ at $f_1(0)=0.07 $on the critical surface in $d=5$.}
\label{fig:13}
\end{figure}

\begin{figure}
\epsfxsize= 12 cm
\centerline{\epsffile{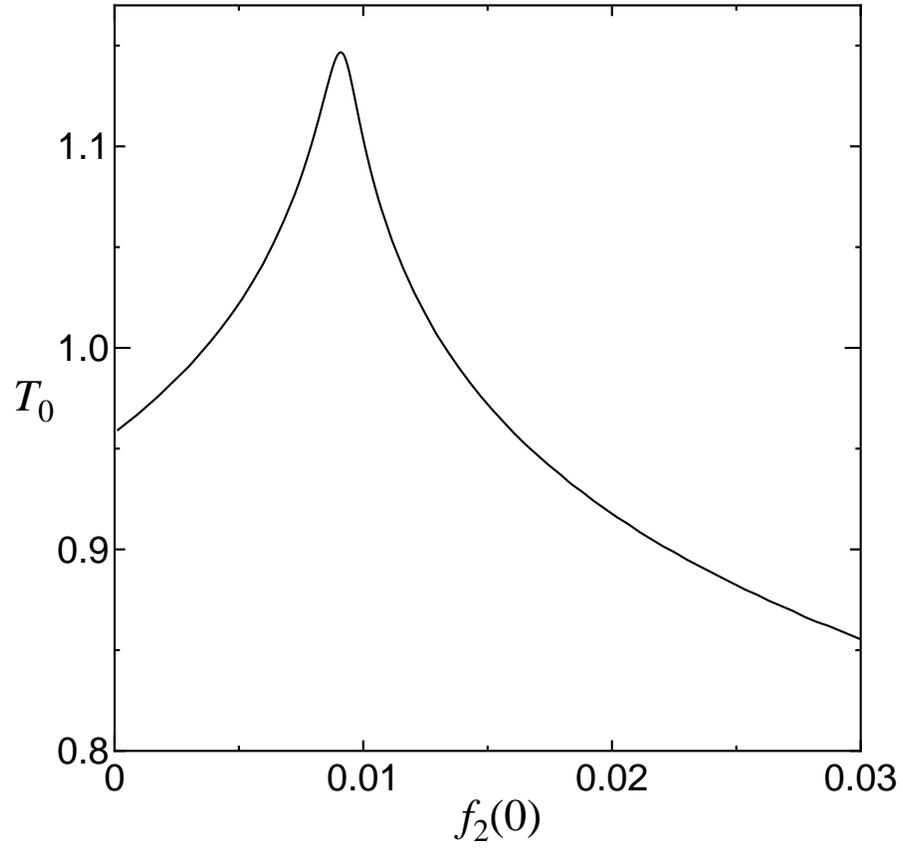}}
\caption{The same as Fig.~13 for $d=6$ with $f_1(0)=0.15 $.}
\label{fig:14}
\end{figure}

\end{document}